\documentclass[number,sort&compress,5p]{elsarticle}

\usepackage{amsmath}
\usepackage{amssymb}
\usepackage{multirow}

\journal{International Journal of Machine Tools and Manufacture}

\renewcommand{\figurename}{Fig.}
\renewcommand{\tablename}{Table}

\begin{document}

\begin{frontmatter}
\title{Evaluation of servo, geometric and dynamic error sources on five axis high-speed machine tool.}

\author[poly,lurpa]{L.~Andolfatto\corref{loic}\footnote{Present address: Laboratoire Universitaire de Recherche en Production Automatis\'ee, \'Ecole Normale Sup\'erieure de Cachan, 61 Avenue du Pr\'esident Wilson, 94230 Cachan, France}}
\ead{loic.andolfatto@lurpa.ens-cachan.fr}

\author[lurpa]{S.~Lavernhe}
\ead{sylvain.lavernhe@lurpa.ens-cachan.fr}

\author[poly]{J.R.R.~Mayer}
\ead{rene.mayer@polymtl.ca}

\cortext[loic]{Corresponding author. {Tel.:+33\,1\,47\,40\,27\,57}; {Fax:+33\,1\,47\,40\,22\,20}}

\address[poly]{Mechanical Engineering Department, \'Ecole Polytechnique de Montr\'eal, PO Box 6079, Station Centre-ville, Montr\'eal, Quebec, Canada H3C 3A7}

\address[lurpa]{Laboratoire Universitaire de Recherche en Production Automatis\'ee, \'Ecole Normale Sup\'erieure de Cachan, 61 Avenue du Pr\'esident Wilson, 94230 Cachan, France}


\begin{keyword}
Machine tool \sep geometric errors \sep servo errors \sep dynamic errors \sep high-speed machining
\end{keyword}

\begin{abstract}
Many sources of errors exist in the manufacturing process of complex shapes. Some approximations occur at each step from the design geometry to the machined part. 

The aim of the paper is to present a method to evaluate the effect of high speed and high dynamic load on volumetric errors at the tool center point.

The interpolator output signals and the machine encoder signals are recorded and compared to evaluate the contouring errors resulting from each axis follow-up error. The machine encoder signals are also compared to the actual tool center point position as recorded with a non-contact measuring instrument called CapBall to evaluate the total geometric errors. The novelty of the work lies in the method that is proposed to decompose the geometric errors in two categories: the quasi-static geometric errors independent from the speed of the trajectory and the dynamic geometric errors, dependent on the programmed feed rate and resulting from the machine structure deflection during the acceleration of its axes. 

The evolution of the respective contributions for contouring errors, quasi-static geometric errors  and dynamic geometric errors is experimentally evaluated and a relation between programmed feed rate and dynamic errors is highlighted.
\end{abstract}
\end{frontmatter}

\section{Introduction}

High speed machining is known to reduce machining time and improve surface quality, due to its particular cutting process \cite{lave:2008}. The ability to control the tool orientation with respect to the workpiece also leads to productivity improvements.

The manufacturing process includes several steps. First, the part to machine, generally described by a digital model, is used to generate the appropriate tool path for machining via a computer-aided machining  software. This step generally comes with approximations in the part geometry: a complex shape can be interpolated and discretized, leading to differences between the generated tool path and the numerical model of the part.

The post processing of the computer-aided machining file is required to convert the tool path into CNC commands. The tool path, originally described in the workpiece coordinate system (WCS) must be transformed in the joint coordinate system (JCS). This inverse kinematic transformation is most often performed during the post processing or sometime directly calculated by the NC-unit, but in any case, approximations may occur \cite{lave:2008,lave-tofo:2008}.

Then, the NC-unit interpolates the given tool path expressed in the JCS to provide synchronous and dynamically admissible input for the machine drives. Each machine axis is subject to follow-up errors, leading to orientation and position errors of the tool \cite{lave:2008}.

The machine structure, in its quasi-static state, suffers from link and motion errors causing position and orientation errors of the tool \cite{abba:2002,knap:2009,zarg:2009}. Moreover, thermal variations of the machine is another source of errors \cite{schm:2008}.

When programmed feed rate and actual velocity increase, as is usual during the high-speed machining process, the dynamic solicitations of the structure become higher because of the acceleration of the different parts of the machine. This may result in some alterations of the machine geometry, causing further tool position and orientation errors \cite{alti:2005,cano:2008}.

Finally, the forces appearing during the cutting process lead to tool deflection, causing another source of machined surface location error \cite{schm:2008}.

Schmitz \emph{et al.} compared the contributions of the geometric, thermal, controller and cutting forces errors to inaccuracies on a three-axis geometry, using different methods and instruments to measure and evaluate the respective contributions at different programmed feed rates \cite{schm:2008}. Zhu \emph{et al.} identified the repeatable errors of machine tool using B-spline models using a cross-grid scale system \cite{Zhu2010}. Ibaraki \emph{et al.} identified kinematic errors of a five axis machine tool from geometric errors of finished workpieces and compared the results to ball bar measurements \cite{Ibaraki2010}. Bringmann and Knapp measured the relative tool center point (TCP) deviations between a sphere and a nest of four linear probes to calibrate a five axis machine tool \cite{knap:2006}. Erkan and Mayer used the machine own touch trigger probe to measure a set of sphere of uncalibrated relative position to calibrate the machine and assess its positioning performance \cite{Erkan2010}. Finally, Schwenke \emph{et al.} provide a comprehensive overview of other measurement method to identify machine tool error in \cite{Schwenke2008}.

A method to evaluate the contributions of error sources to the Cartesian volumetric errors at the tool center point of a five axis machine tool at different programmed feed rates is described in this paper. The Cartesian volumetric errors are defined as the three components of the vector from the theoretical position of the TCP relative to the workpiece frame to its actual position. They are decomposed into the contribution of three error sources:
\begin{itemize}
\item the effect of follow-up errors of the axis drives, later called \emph{contouring errors} and written $\boldsymbol{\delta}_c$;
\item the quasi-static geometric errors of the machine, which include link and motion errors and thermal drift, written $\boldsymbol{\delta}_{qs}$;
\item the dynamic geometric errors, resulting from the machine structure deflection under dynamic load, and written $\boldsymbol{\delta}_d$.
\end{itemize}

The evaluation of those three error sources can help quantifying the relative impact of dynamic errors to the total volumetric errors, and the relevance of an associated model for corrective actions. 

The paper begins with a presentation of the principle of the method is presented. Then, the mathematical models and the hypothesis drawn to evaluate each error source is developed. The experimental setup is described and finally, the results gathered from low to high speed motion are given and discussed.

\section{Principle of the method}

\begin{figure}[tbh]
\center
\includegraphics[scale=1]{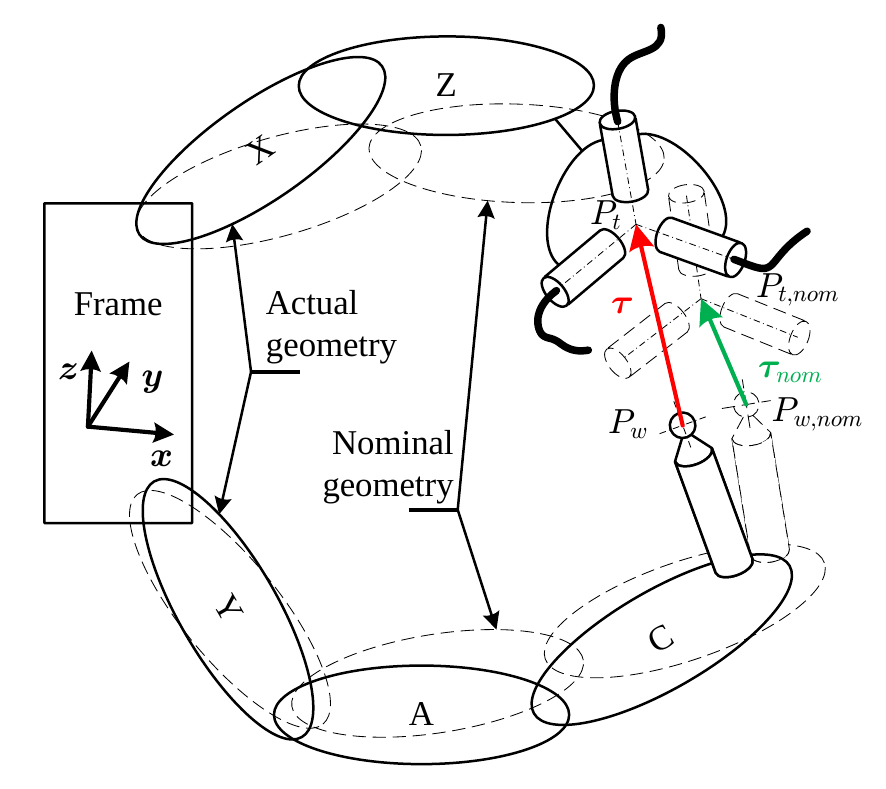}
\caption{Difference between the nominal position of the TCP relatively to workpiece $\boldsymbol{\tau}_{nom}$ calculated from controller inputs and direct kinematic model and its actual position $\boldsymbol{\tau}$ measured with the CapBall.}
\label{fig:figure_machine}
\end{figure}

The nominal position of the tool center point relative to the workpiece is evaluated with the nominal direct kinematic transformation (DKT) of the machine fed with the controller inputs of the machine.

The actual position of the tool relative to the workpiece is measured with a non-contact sensor called CapBall, developed in house \cite{zarg:2009}, which provides the position of a ball mounted on the machine table relative to the TCP.

The difference between the nominal and the actual position of the tool relative to the workpiece, pictured on \figurename~\ref{fig:figure_machine}, is due to the servo-errors of each axis of the machine and to all the geometric defects of the machine structure. This difference corresponds to the sum of $\boldsymbol{\delta}_c$, $\boldsymbol{\delta}_{qs}$ and $\boldsymbol{\delta}_d$, respectively the contouring error, the quasi-static geometric error and the dynamic geometric error.

The contouring error $\boldsymbol{\delta}_c$ can be evaluated by recording the machine controller inputs and encoder actual values\footnote{From position feedback linear encoders for linear axes and angular position feedback encoders for rotary axes.} and comparing the relative position of the tool and the workpiece from the two points of view.

At low velocity, the dynamic geometric errors $\boldsymbol{\delta}_d$ can be considered negligible because the dynamic load is low with respect to the usual high stiffness of machines. Thus, once contouring errors have been accounted for, the remaining errors measured at a low programmed feed rate can be defined as the quasi-static geometric errors $\boldsymbol{\delta}_{qs}$.

Finally when the programmed feed rate and the machine velocity increase, alterations of the machine structure under dynamic load cause the dynamic geometric errors $\boldsymbol{\delta}_d$. They can be evaluated by measuring the total geometric errors along a single trajectory at different programmed feed rates and removing the contouring error and quasi-static geometric error contributions.

\begin{figure*}
\center
\includegraphics[width=\linewidth]{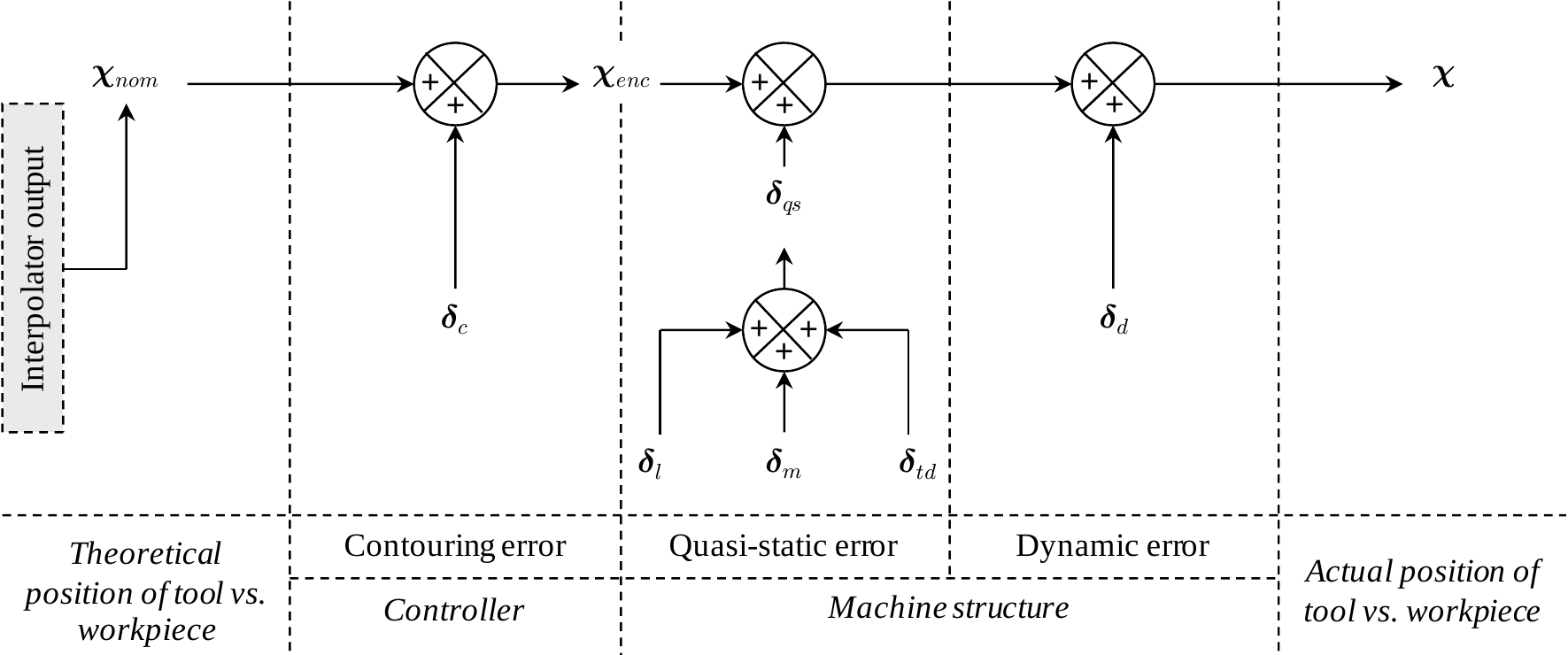}
\caption{Decomposition of the total error into contouring error (subsection \ref{sec:contouring}), quasi-static geometric error (subsection \ref{sec:qs}) and dynamic geometric error (subsection \ref{sec:dynamic}).}
\label{fig:figure_principe}
\end{figure*}

\section{Model for error contributions}

\subsection{Decomposition of the total errors}

During a measurement sequence, $n$ points, equally distributed, are recorded on each of the three channels of the CapBall.

The CapBall readings provide the values of the 3 components of the measured difference between $P_t$ -- the actual tool center point -- and $P_w$ -- the actual center of the master ball mounted on the machine table --, written $\boldsymbol{\tau}$ for each of the $n$ recorded points (see \figurename~\ref{fig:figure_machine}):

\begin{equation}
\boldsymbol{\tau} = P_t - P_w 
\end{equation}

The results are gathered in a $n \times 3$ matrix written $\boldsymbol{\chi}$, as shown in eq.\eqref{eq:delta_chi}:

\begin{equation}
\boldsymbol{\chi}=\begin{pmatrix} {\boldsymbol{\tau}_1}^T \\ \vdots \\ {\boldsymbol{\tau}_n}^T \end{pmatrix}
\label{eq:delta_chi}
\end{equation}

\noindent
where $\boldsymbol{\tau}_k$ is the CapBall reading for $k$-th point of the trajectory. The effect of the master ball and sensing head set-up positioning errors on volumetric errors are previously identified with the method described in \cite{zarg:2009,ando:2011} and removed from $\boldsymbol{\chi}$.

The direct kinematic transformation (DKT) associated with the nominal machine geometry and the five axis controller inputs allows to calculate:
\begin{itemize}
\item the position of $P_{t,nom}$ -- the nominal TCP -- relative to the machine frame, considering the machine at its nominal geometry;
\item the position of  $P_{w,nom}$ -- the nominal master ball center -- also relative to the machine frame and also considering the machine with no geometric defect.
\end{itemize} 

The difference of position between those two points can be expressed as $\boldsymbol{\tau}_{nom}$ calculated with the axis controller inputs (see \figurename~\ref{fig:figure_machine} and eq.\eqref{eq:def_tau_nom}).

\begin{equation}
\boldsymbol{\tau}_{nom} = P_{t,nom} - P_{w,nom}
\label{eq:def_tau_nom}
\end{equation}

\begin{equation}
\boldsymbol{\tau}_{nom} = DKT(C_{nom},A_{nom},Y_{nom},X_{nom},Z_{nom}) 
\label{eq:def_tau_nom_dkt}
\end{equation}

The nominal difference matrix $\boldsymbol{\chi}_{nom}$ -- calculated from the controller inputs -- is formed as described in eq.\eqref{eq:chi_nom}.

\begin{equation}
\boldsymbol{\chi}_{nom}=\begin{pmatrix} {\boldsymbol{\tau}_{nom,1}}^T \\ \vdots \\ {\boldsymbol{\tau}_{nom,n}}^T \end{pmatrix}
\label{eq:chi_nom}
\end{equation}

Then, the difference between $\boldsymbol{\chi}$ and $\boldsymbol{\chi}_{nom}$ expresses the effect of the contouring error, the quasi-static geometric error and the dynamic geometric error, as explicited in \figurename~\ref{fig:figure_principe} and eq.\eqref{eq:contributions_delta_chi}:

\begin{equation}
\boldsymbol{\chi} - \boldsymbol{\chi}_{nom} = \boldsymbol{\delta}_c +  \boldsymbol{\delta}_{qs} + \boldsymbol{\delta}_d
\label{eq:contributions_delta_chi}
\end{equation}

\subsection{Contouring error}
\label{sec:contouring}

\begin{figure*}[tbh]
\center
\includegraphics[width=\linewidth]{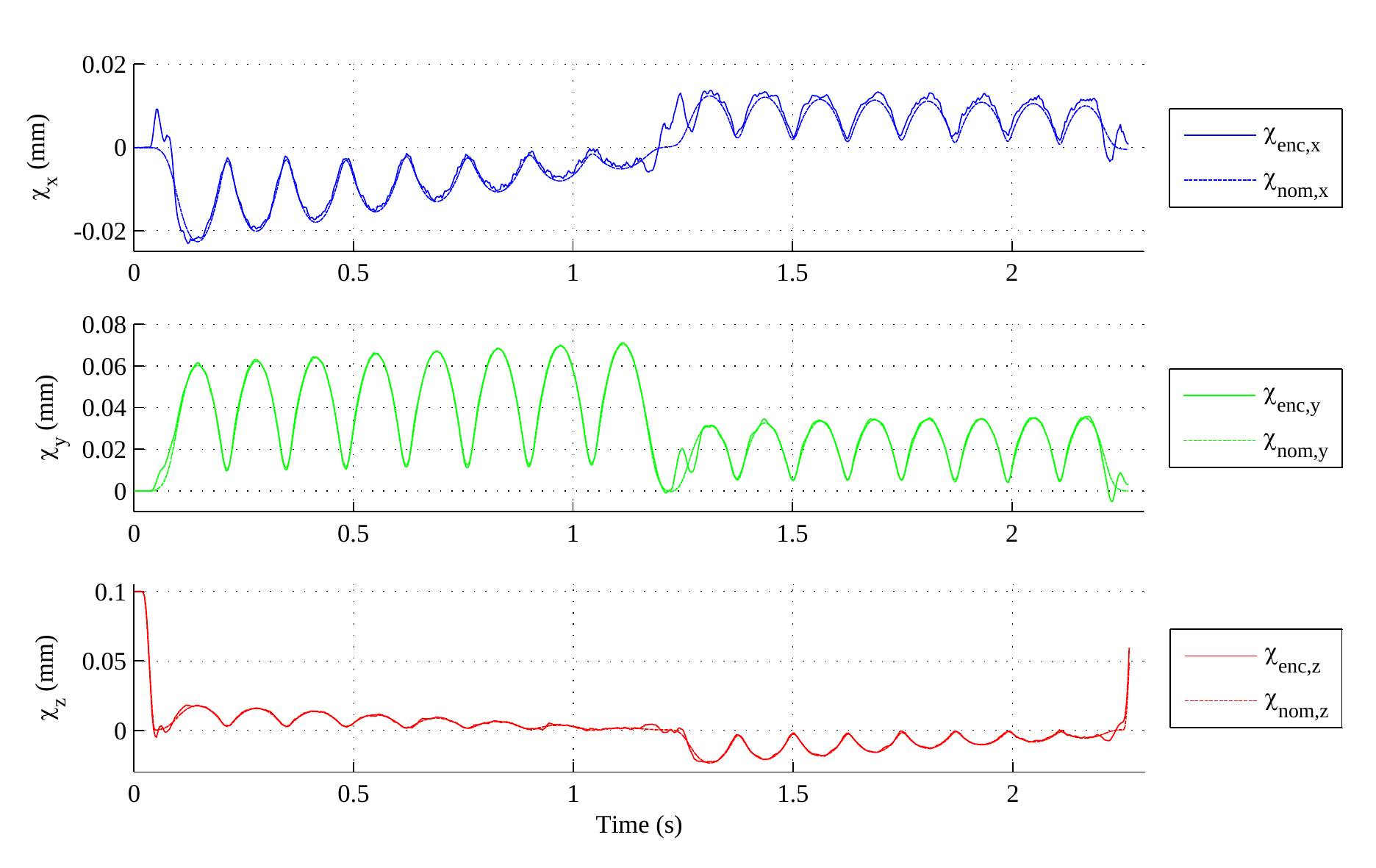}
\caption{Example of the three components of the Cartesian volumetric difference projected in the machine frame, computed with synchronised controllers inputs ($\boldsymbol{\chi_{nom}}$) and actual encoders values  ($\boldsymbol{\chi_{enc}}$) for a programmed feed rate $F=5,000\,mm/min$.}
\label{fig:synchro_delta_f}
\end{figure*}

\begin{figure}[tb]
\center
\includegraphics[scale=1]{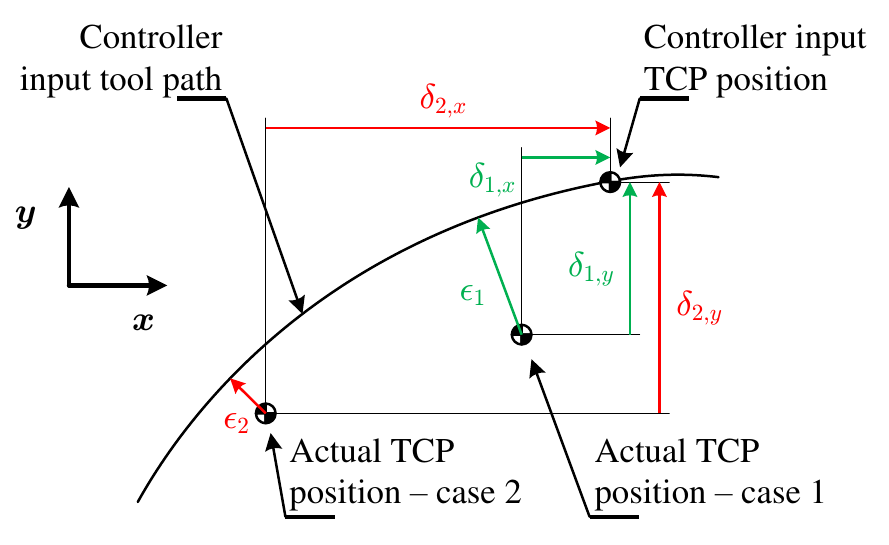}
\caption{Schematic example of the difference between axis by axis Cartesian follow-up error $\boldsymbol{\delta}$ and contouring error $\boldsymbol{\epsilon}$ (adapted from \cite{senc:2009-1}).}
\label{fig:figure_contouring}
\end{figure}

The actual effect of follow-up errors on Cartesian volumetric errors is defined by the \emph{contouring error} $\epsilon$, which is the orthogonal deviation of the actual tool path with respect to the controller inputs tool path in the part coordinate system \cite{senc:2009-1}. The difference between Cartesian follow-up error and contouring error is schematically depicted in \figurename~\ref{fig:figure_contouring}: in case 1, the follow-up errors along $x$ and $y$ are smaller than in case 2, but the resulting contouring error $\epsilon_1$ is higher than $\epsilon_2$.

The difference $\boldsymbol{\tau}_{enc}$ is calculated for each point, considering the direct kinematic transform of the machine at its nominal geometry and the encoder values:

\begin{equation}
\boldsymbol{\tau}_{enc} = P_{t,enc} - P_{w,enc}
\label{eq:def_tau_enc}
\end{equation}

\begin{equation}
\boldsymbol{\tau}_{enc} = DKT(C_{enc},A_{enc},Y_{enc},X_{enc},Z_{enc}) 
\label{eq:def_tau_enc_dkt}
\end{equation}

The encoder actual values volumetric difference matrix is built by the same process as in eq.\eqref{eq:chi_nom}:

\begin{equation}
\boldsymbol{\chi}_{enc}=\begin{pmatrix} {\boldsymbol{\tau}_{enc,1}}^T \\ \vdots \\ {\boldsymbol{\tau}_{enc,n}}^T \end{pmatrix}
\label{eq:chi_enc}
\end{equation}

The contouring error is evaluated by computing the difference between the nominal volumetric difference matrix $\boldsymbol{\chi}_{nom}$ and the encoder actual values volumetric difference matrix $\boldsymbol{\chi}_{enc}$.

The controller input signals must be synchronised with the encoders actual values to evaluate the effect of the follow-up errors on the volumetric errors at the tool tip without taking into account the delay introduced by the recording process of the axis controllers signals. The recordings performed show that the delay between encoder signals and controller inputs signals was $18\,ms$ regardless of the programmed feed rate $F$. The \emph{encoder volumetric difference matrix} $\boldsymbol{\chi}_{enc}$ is obtained by bringing backward the encoder values by $18\,ms$. The synchronised controller inputs and encoders volumetric errors are plotted on \figurename~\ref{fig:synchro_delta_f}.\\

The contouring error along the experimental trajectory is given by eq.\eqref{eq:delta_f}:

\begin{equation}
\boldsymbol{\delta}_c = \boldsymbol{\chi}_{enc} - \boldsymbol{\chi}_{nom}
\label{eq:delta_f}
\end{equation}

\noindent
where $\boldsymbol{\delta}_c$ has a $n \times 3$ size, representing the contouring error expressed in the machine frame.

Combining eqs.\eqref{eq:contributions_delta_chi}  and \eqref{eq:delta_f} allows to express the quasi-static geometric error and the dynamic geometric error as the difference of the recorded signals using eq.\eqref{eq:contrib_geo}:

\begin{equation}
\boldsymbol{\chi} - \boldsymbol{\chi}_{enc} =  \boldsymbol{\delta}_{qs} + \boldsymbol{\delta}_d
\label{eq:contrib_geo}
\end{equation}

\subsection{Quasi-static geometric errors}
\label{sec:qs}
\subsubsection{Definition}

The \emph{quasi-static} geometric error $\boldsymbol{\delta}_{qs}$ is defined as the geometric error independent of the machine velocity. It is decomposed into three sources:

\begin{itemize}
\item the link errors that represents the axis to axis location errors of the machine;
\item the motion errors of each axis;
\item the thermal drift of the machine.
\end{itemize}

Consequently, the quasi-static geometric error is written as:
\begin{equation}
\boldsymbol{\delta}_{qs} = \boldsymbol{\delta}_l + \boldsymbol{\delta}_m + \boldsymbol{\delta}_{td}
\label{eq:delta_{qs}}
\end{equation}
\noindent
where $\boldsymbol{\delta}_l$, $\boldsymbol{\delta}_m$ and $\boldsymbol{\delta}_{td}$ are respectively the contribution of the link errors, the motion errors and the thermal drift.

\subsubsection{Link errors}

\begin{figure}[tb]
\center
\includegraphics[scale=1]{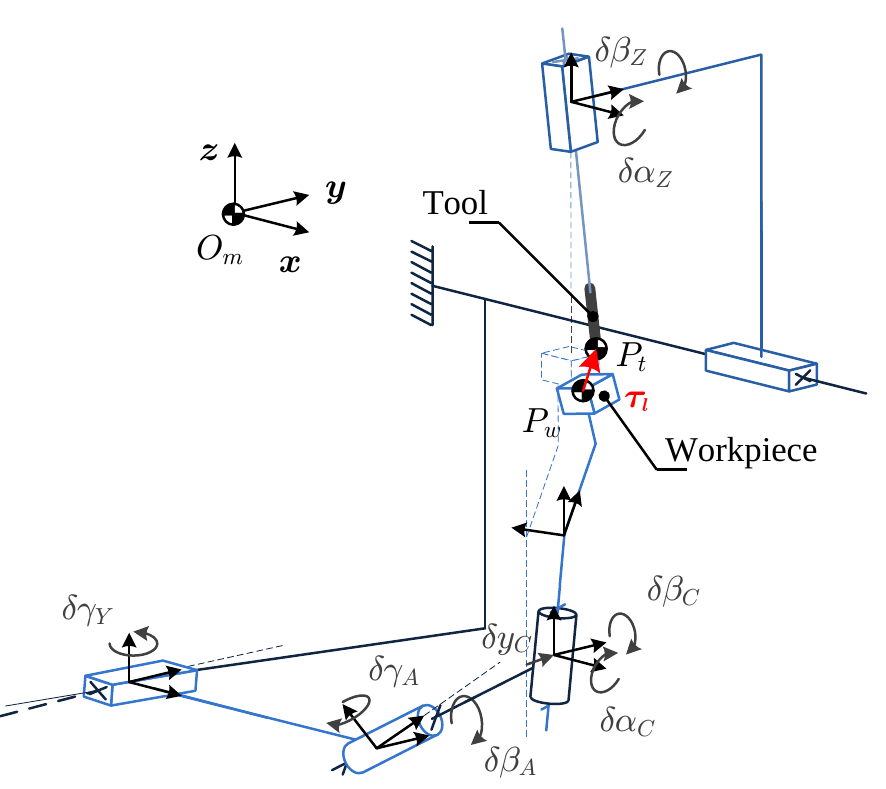}
\caption{Influence of the eight link errors on the geometric errors for the five axis machine tool studied.}
\label{fig:schema_defauts_machine}
\end{figure}

Mir \emph{et al.} proposed a model to express the effect of axis-to-axis link errors on volumetric errors \cite{abba:2002}. In this model, the link errors for a five axis machine tool are described by a set of eight geometric errors of the machine structure. For the example in \figurename~\ref{fig:schema_defauts_machine}, the link errors are:
\begin{itemize}
\item $\delta \gamma_Y$, the out-of-squareness between X and Y;
\item $\delta \alpha_Z$, the out-of-squareness between Y and Z;
\item $\delta \beta_Z$, the out-of-squareness between X and Z;
\item $\delta \beta_A$, the tilt of A around $y$;
\item $\delta \gamma_A$, the tilt of A around $z$;
\item $\delta \alpha_C$, the tilt of C around $x$;
\item $\delta \beta_C$, the tilt of C around $y$;
\item $\delta y_C$, the offset of A relative to C in $y$.
\end{itemize}
\noindent
They are gathered in an array written $\boldsymbol{\delta q}_l$:

\begin{equation}
\boldsymbol{\delta q}_{l} = \left( \delta \gamma_Y \, \delta \alpha_Z \, \delta \beta_Z \, \delta \beta_A \, \delta \gamma_A \, \delta \alpha_C \, \delta \beta_C \, \delta y_C \right)^T
\label{eq:delta q_{l}}
\end{equation}

The model is based on a Jacobian principle. The volumetric errors due to the eight link errors of the machine at a pose $k$, written as a three components column vector $\boldsymbol{\tau}_{l,k}$, is given by eq.\eqref{eq:machine_error}:

\begin{equation}
\boldsymbol{\tau}_{l,k} = \boldsymbol{J}_{l,k} \cdot \boldsymbol{\delta q}_l
\label{eq:machine_error}
\end{equation}

\noindent
where $\boldsymbol{J}_{l,k}$ is the link errors Jacobian matrix expressed at a pose $k$.  More details are available in \cite{abba:2002,zarg:2009}. Zargarbashi and Mayer described a method using CapBall data to identify the eight machine errors \cite{zarg:2009,ando:2011}. This method is also used here to identify the eight parameters of the model.

Then, for each known pose of the machine, corresponding to the encoder actual values, the link errors Jacobian matrix $\boldsymbol{J}_{l,k}$ and then $\boldsymbol{\tau}_{l,k}$ are calculated, leading to the expression for the contributions of the link errors $\boldsymbol{\delta}_l$ (eq.\eqref{eq:def_delta_chi_l}):
 
\begin{equation}\boldsymbol{\delta}_l = 
\begin{pmatrix} {\boldsymbol{\tau}_{l,1}}^T \\ \vdots \\  {\boldsymbol{\tau}_{l,n}}^T \end{pmatrix}
\label{eq:def_delta_chi_l}
\end{equation} 

\noindent
where $\boldsymbol{\delta}_l$ has a $n \times 3$ size, representing the three Cartesian components of the modelled contribution of the link errors to the volumetric errors at each point along the trajectory.

\subsubsection{Motion errors}

\begin{figure*}[tbh]
\center
\includegraphics[width=\linewidth]{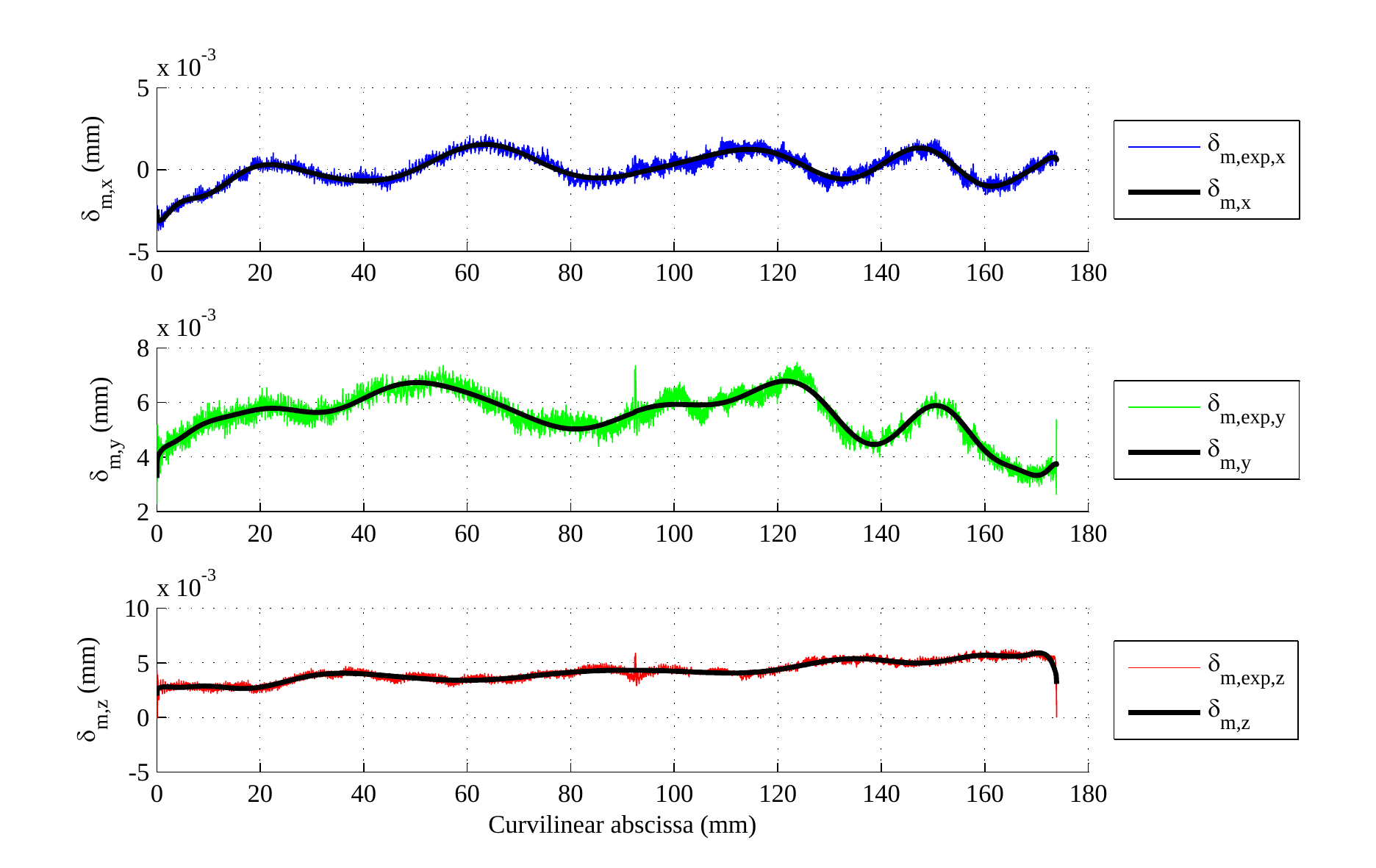}
\caption{Comparison of the local polynomial model for motion errors contributions $\boldsymbol{\delta}_{m}$ and measured residual errors $\boldsymbol{\delta}_{m,exp}$ (sum of the motion errors and local dynamic errors) along a trajectory at a programmed feed rate of $F=1\,000\,mm/min$.}
\label{fig:delta_chi_p}
\end{figure*}

A reference thermal state of the machine, in which the value of the drift is considered to be zero, must be chosen for a test at low programmed feed rate. Then, the portion of measured errors during this test not explained by the link errors is calculated by eq.\eqref{eq:residus} and is attributed to the motion errors and local dynamic errors.

\begin{equation}
\left\{ \begin{array}{l}
\boldsymbol{\chi} -  \boldsymbol{\chi}_{enc} - \boldsymbol{\delta}_l = \boldsymbol{\delta}_m + \boldsymbol{\delta}_d\\
\boldsymbol{\delta}_{td} = 0 \end{array} \right.
\label{eq:residus}
\end{equation}

In \cite{knap:2009}, Bringmann and Knapp proposed a Fourier series to model the motion errors of a link. Slamani \emph{et al.} proposed a polynomial model for motion errors in \cite{slam:2009}. In both cases the effects of motion errors are defined by continuous and smooth mathematical functions.

In order to evaluate the volumetric contribution of motion errors, a polynomial model is used. The only aim of this model is to locally predict the effect of the five axes motion errors on volumetric errors along the experimental trajectory to subtract it from the measured volumetric errors. The polynomial method proved to be easier to implement while providing satisfactory results.

The dynamic geometric error proved to be negligible at a low programmed feed rate (compared to link and motion errors), and they are visible only locally\footnote{Dynamic geometric error mainly appears at the beginning and the end of a trajectory, and at sharp corners.} along the trajectory, as depicted in \figurename~\ref{fig:delta_chi_p}.

A normalised path parameter $\boldsymbol{t}_n$ is created to describe the experimental trajectory. It has $n$ linearly distributed components, from 0 for the first point to 1 for the last point. The residual errors matrix is curve-fitted with polynomial functions of $\boldsymbol{t}_n$. The approximation leads to 3 identified polynomials $\boldsymbol{P}_x$,  $\boldsymbol{P}_y$ and  $\boldsymbol{P}_z$. The applied polynomial curve fitting leads to a filtering of local dynamic errors and measurement noise.

Experimental results show that, for the machine tested, 20 was a suitable value for the degree of the polynomial. Below 20, the curve fitting did not represent the residual errors well, and above 20, some unexpected oscillations appeared. The model predicted contribution of motion errors is compared to actual measurement in \figurename~\ref{fig:delta_chi_p}.

The contributions of motion errors to volumetric errors $\boldsymbol{\delta}_m$ are modelled in eq.\eqref{eq:delta_chi_m}:

\begin{equation}
\boldsymbol{\delta}_m=
\begin{pmatrix}
\boldsymbol{P}_x(0) 	& \boldsymbol{P}_y(0) 	& \boldsymbol{P}_z(0) 	\\
\vdots	&	\vdots	& \vdots \\
\boldsymbol{P}_x(t_{n,k}) 	& \boldsymbol{P}_y(t_{n,k}) 	& \boldsymbol{P}_z(t_{n,k}) 	\\
\vdots	&	\vdots	& \vdots \\
\boldsymbol{P}_x(1) 	& \boldsymbol{P}_y(1) 	& \boldsymbol{P}_z(1) 
\end{pmatrix}
\label{eq:delta_chi_m}
\end{equation}

\noindent
where $t_{n,k}$ is the $k$-th component of the normalised path parameter.

The main weakness of this method is to compute motion error contributions based on the time instead of the actual pose of the machine at each point. However, the small differential rate of the contributions of the motion errors reduces the effect of the approximation used.

\subsubsection{Thermal drift}

Even if a warm-up is performed and care is taken to favour the thermal stability of the machine, a small thermal drift can be present.

As the execution of an experimental trajectory can take less than $15 \,s$, it can be considered as instantaneous compared to the movement of the sensing head due to thermal drift. From that point of view, the effect of the thermal drift on the volumetric errors can be modelled as an offset from the reference thermal state on the measured volumetric errors. For each test, this offset is given by the mean value of the difference between the measured deviation $\boldsymbol{\chi}$ and the encoder only prediction with a nominal model, the modelled link errors and the motion errors contributions, as expressed in eq.\eqref{eq:offset}:

\begin{equation}
\boldsymbol{td} = \frac{1}{n} \sum_{i=1}^n \left[ \left( ^i\boldsymbol{\chi} - ^i\boldsymbol{\chi}_{enc} \right) - ^i\boldsymbol{\delta}_l - ^i\boldsymbol{\delta}_m \right]
\label{eq:offset}
\end{equation}

\noindent
where $^i\boldsymbol{\chi}$ is the $i$-th line of the matrix $\boldsymbol{\chi}$.

Finally, the drift, defined as constant for all the measurement point of a test is expressed by a $n \times 3$ matrix $\boldsymbol{\delta}_{td}$, composed of $n$ lines equal to $\boldsymbol{td}$:

\begin{equation}
\boldsymbol{\delta}_{td} = \begin{pmatrix}\boldsymbol{td} \\ \vdots \\ \boldsymbol{td} \end{pmatrix}_{(n \times 3)}
\label{eq:delta_td}
\end{equation}

\subsection{Dynamic geometric errors}
\label{sec:dynamic}
The \emph{dynamic} geometric errors are here defined as the additional errors occurring when programmed feed rate, and so dynamic forces on the machine structure, increases. The dynamic errors result from varying alterations of the machine components under dynamic forces such as deflection of the machine structure.

Eqs.\eqref{eq:contributions_delta_chi} and \eqref{eq:delta_{qs}} allow to express the dynamic geometric errors $\boldsymbol{\delta}_d$ as the contribution that does not show at low speed, in eq.\eqref{eq:delta_d}:

\begin{equation}
\boldsymbol{\delta}_d =  \left( \boldsymbol{\chi} - \boldsymbol{\chi}_{enc} \right) - \boldsymbol{\delta}_{qs} 
\label{eq:delta_d}
\end{equation}

\section{Experimental setup and conditions}
\label{par:experimental}

\subsection{Machine tool}

\begin{figure}[tbh]
\center
\includegraphics[scale=1]{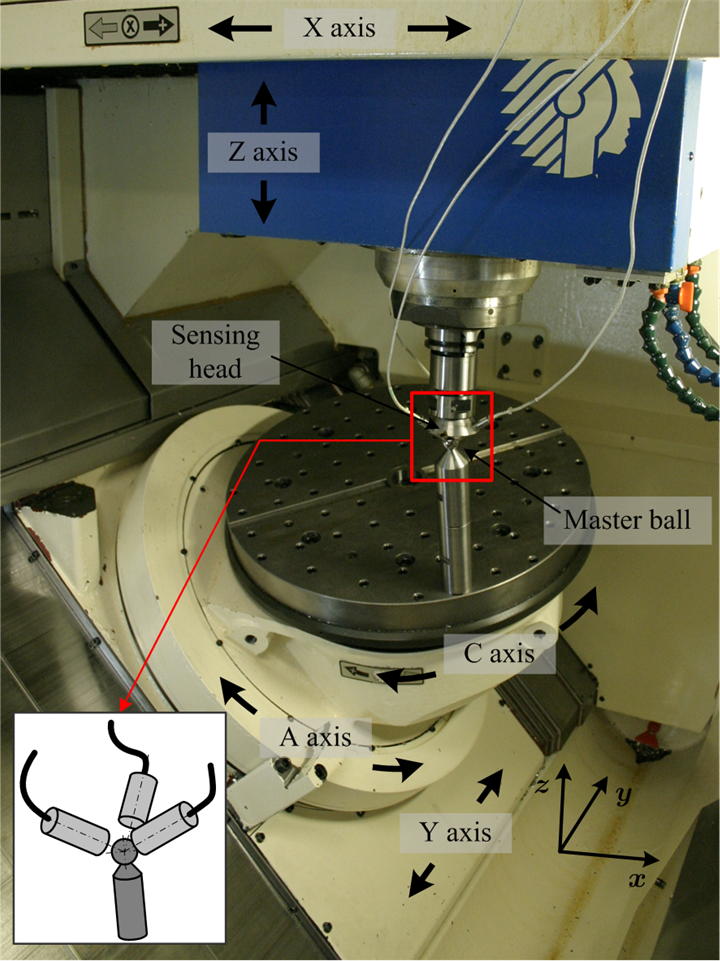}
\caption{Huron KX8-five five axis machine tool fitted with the modified CapBall system.}
\label{vue_machine}
\end{figure}

\begin{figure*}[tbh]
\center
\includegraphics[width=\linewidth]{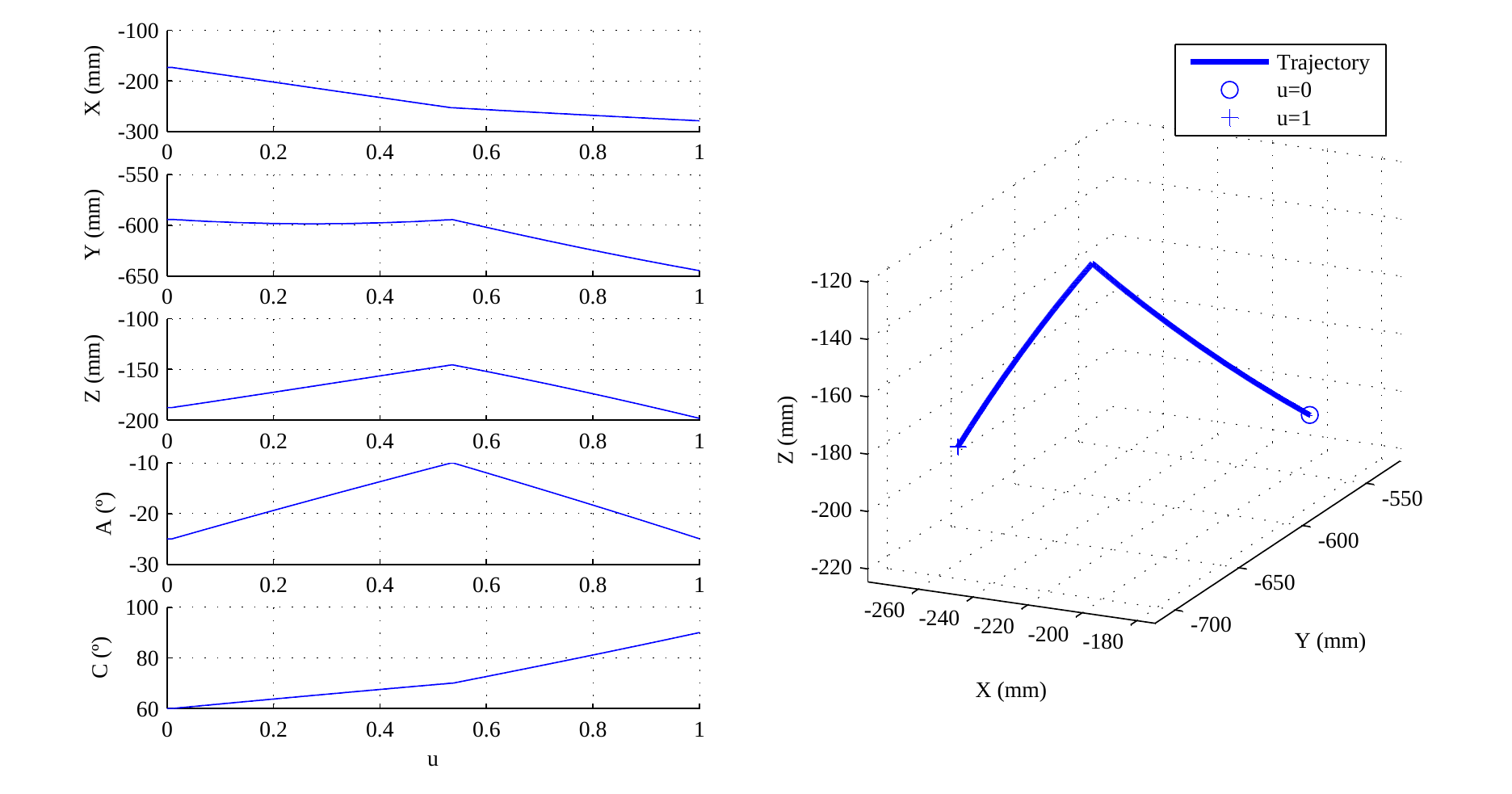}
\caption{Experimental trajectory: left, evolution of each axis setpoint as a function of a normalised path parameter $u$ representing the linearly scaled curvilinear abscissa and right, view of the master ball center $P_w$ trajectory in the machine coordinate system (attached to the Y-stage).}
\label{trajectoire_5x}
\end{figure*}

This study was carried out on a Huron KX8-five five axis machine tool depicted in \figurename~\ref{vue_machine}. This machine has a WCAYFXZT structure, with a 45-degree-tilted A rotary axis. It is equipped with a Siemens Sinumerik 840D Powerline numerical command unit. The NC unit allows the recording of system variables such as axis controller inputs and encoder actual values for each axis (X,Y,Z,A and C). The machine documentation specifies that both axis controller inputs and encoder actual values already include backlash compensations \cite{doconweb-variables}, so this potential source of error is not included in this study.

Previous tests show that the axis controller inputs remain the same for each execution of a program. Thus, the recording of controller inputs can be performed separately.

The encoder actual position and the CapBall measurements are recorded together during the execution of trajectory .

Finally, all 10 channels (5 for the controller input and 5 for the encoder actual values) are gathered.

The NC cycle time is $3\,ms$ and the sampling frequency of the CapBall is $10\,kHz$. To accurately synchronise and compare all the signals, the 10 NC signals are linearly interpolated at $10\,kHz$. A point is added at the beginning of the experimental trajectory to provide a tag dedicated to synchronisation of all signals together. More details are given at the end of subsection \ref{par:trajectory}.

\subsection{External measuring instrument}

The CapBall consists of a precision master ball mounted on the table and a sensing head fitted with three capacitive sensors, mounted in a tool holder in the spindle (see \figurename~\ref{vue_machine}).

The three capacitive sensors axes are nominally orthogonal to each other and intersecting in one point called $P_t$, chosen as the virtual TCP. This forms an orthogonal measurement frame.

The capacitive sensors are pre-calibrated for the spherical target using a set of three high precision linear stages. The response non-linearity is kept under $0.5\%$ as long as the eccentricity is kept within $\pm 300 \, \mu m$ and the distance between the sensor and the ball is in a $\pm 300 \, \mu m$ range centred on the position at which the sensor gives a 0 Volt signal. The machine is programmed to keep the center of the master ball $P_w$ close to $P_t$ to avoid exceeding the measurement range of the sensors. In this range, standard type A uncertainty on the sensors output has been evaluated below $0.4\,\mu m$ \cite{ando:2011}.

The spindle orientation is regulated to keep the sensing head orientation constant relative to the machine coordinate system -- or machine frame -- $(\boldsymbol{x},\boldsymbol{y},\boldsymbol{z})$ (\figurename~\ref{vue_machine}).

To express the position of $P_t$ relative to $P_w$ in the machine tool frame, the orientation of the sensors axes are calibrated. Then, the position of $P_t$ relatively to $P_w$ can be projected in the machine tool frame, to express $\boldsymbol{\tau}$.

The sensing head described in \cite{zarg:2009} was completely re-designed to significantly increase its stiffness and to ensure negligible displacement of the sensors during a measurement under dynamic solicitations in the maximum operating range of the machine tool. The master ball base is a $40\,mm$ diameter steel bar screwed to the machine table via a solid post (\figurename~\ref{vue_machine}).

The sensors readings are acquired trough a LABView application. All the computation are implemented in MATLAB programs.

\subsection{Experimental trajectory}

\label{par:trajectory}

To evaluate the contributions of the contouring, quasi-static and dynamic errors at different feed rates, an experimental trajectory is chosen to provide several characteristics.

It must not last longer than 15 seconds when executed at the lowest programmed feed rate (\emph{i.e.} $1000\,mm/min$),  in order to avoid exceeding the NC unit recording buffer size. This constraint reduces the maximum length of the trajectory, forbidding performing standard tests as Schmitz \emph{et al.} did in \cite{schm:2008}.

The motion of the master ball results from the rotary axis motion so its path is a complex curve. As the sensing head motion was programmed with linear interpolation, a resulting programmed difference of position between $P_w$ and $P_t$ appears. The maximum admissible difference was chosen as $100 \, \mu m$ in each direction of the machine frame. 

\begin{figure*}[tbh]
\center
\includegraphics[width=\linewidth]{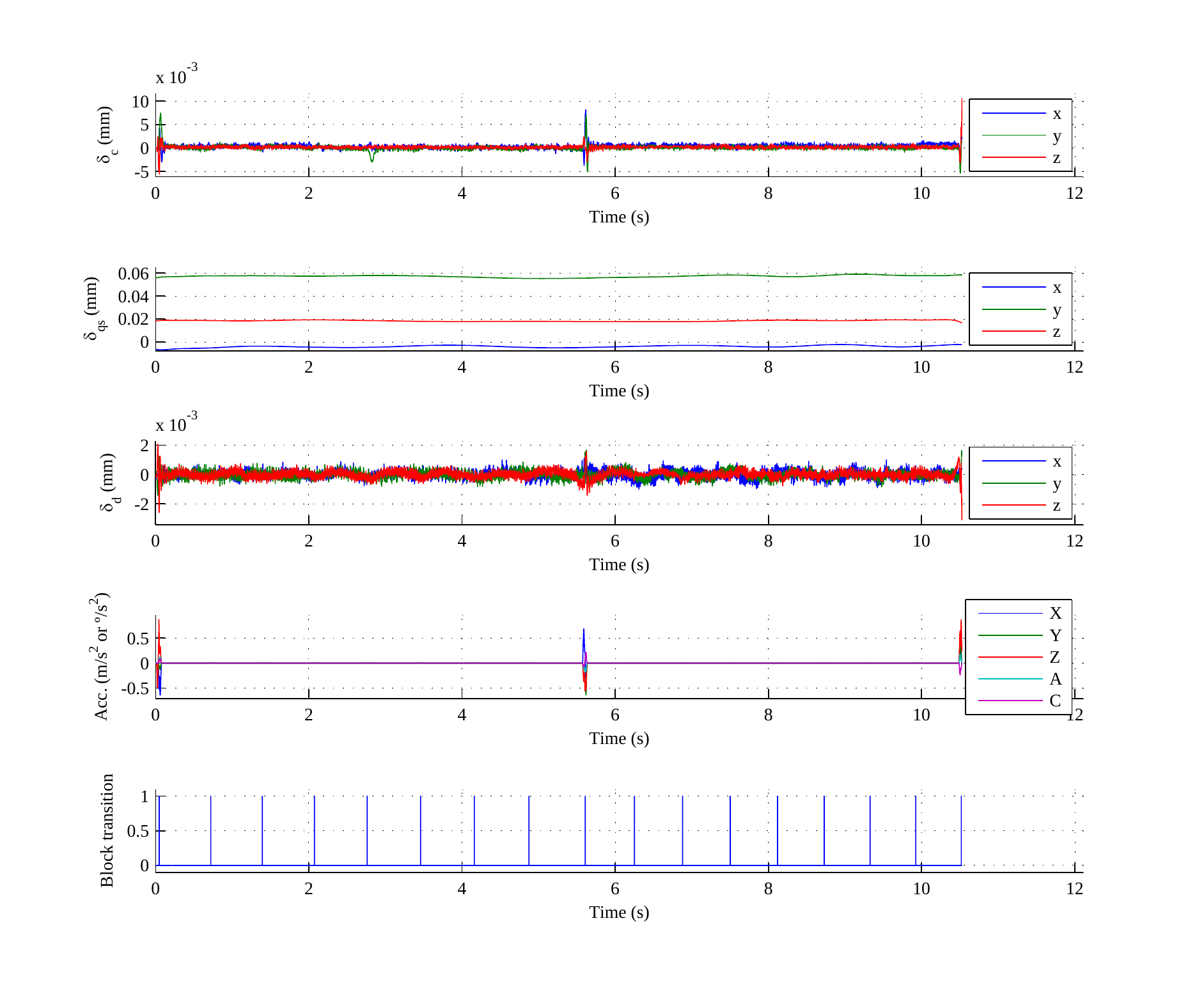}
\caption{Computed contributions of the follow-up errors, quasi-static and dynamic errors and the associated machine axes acceleration and block transition at a programmed feed rate of $F=1\,000\,mm/min$.}
\label{fig:essai1}
\end{figure*}

\begin{figure*}[tbh]
\center
\includegraphics[width=\linewidth]{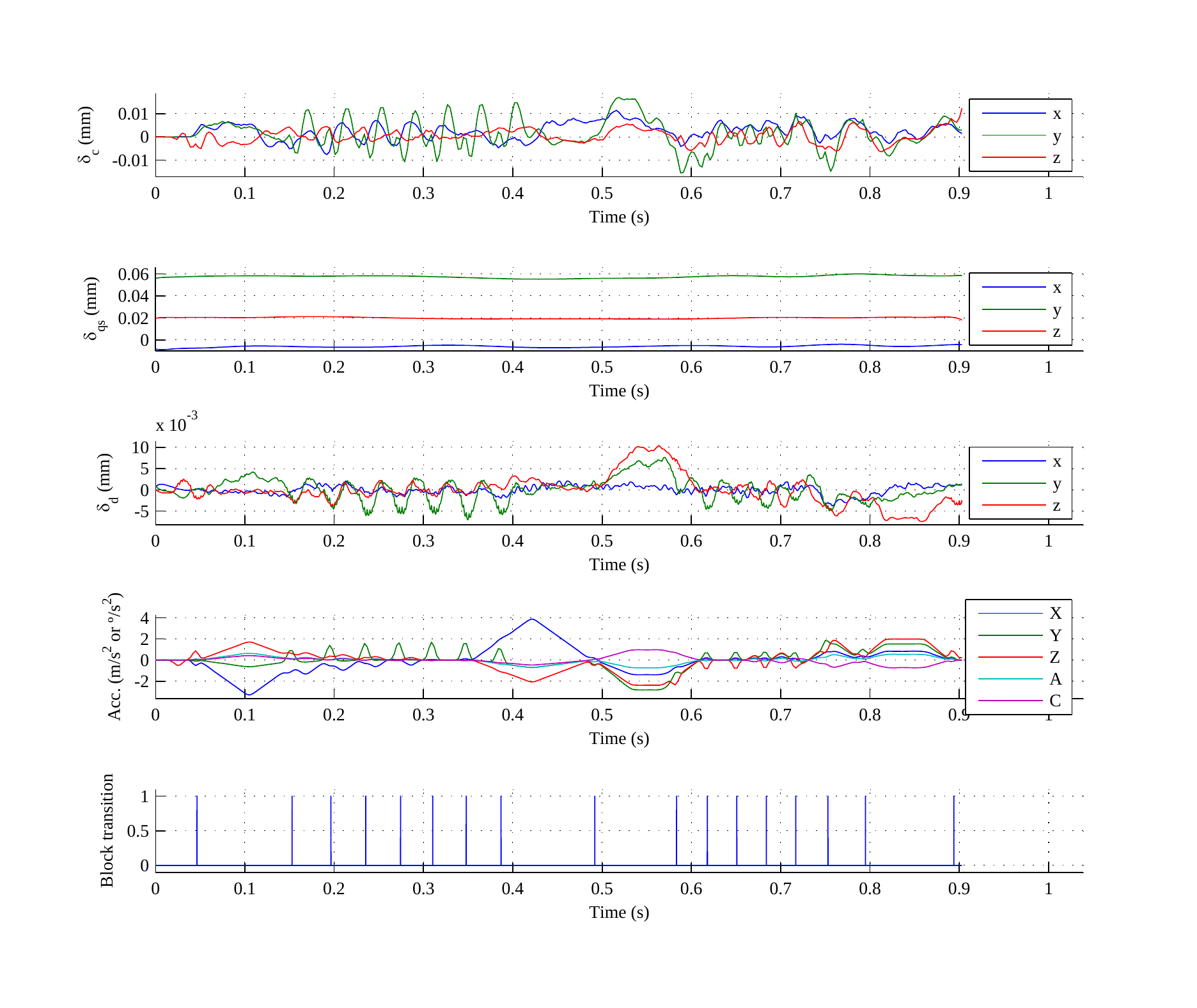}
\caption{Computed contributions of the follow-up errors, quasi-static and dynamic errors and the associated machine axes acceleration and block transition at a programmed feed rate of $F=18\,896\,mm/min$.}
\label{fig:essai6}
\end{figure*}

The experimental trajectory -- with a nearly square corner -- was chosen to provide sizeable geometric discontinuities which require accelerations of several axes. This is the solution chosen to generate a large range of dynamic loads while the programmed feed rate varies. The chosen experimental trajectory is described for each axis as a function of a normalised parameter $u$ varying from $0$ to $1$ in \figurename~\ref{trajectoire_5x}. This parameter represents the linearly scaled curvilinear abscissa along the trajectory. The trajectory, as perceived in the coordinate system attached to the Y-stage, is also depicted. The trajectory is described by 17 linear segments (programmed with \texttt{G01} blocks). The linear interpolation format leads to acceleration peaks at the transition between each segment, as shown in \figurename~\ref{fig:essai6}.

The motion of each axis is programmed with linear interpolation, and all the axes are synchronised (\texttt{FGROUP(X, Y, Z, A, C)}). The machine is commanded in joint space (\texttt{TRAFOOF} option)to provide the sought dynamic solicitations. To avoid critical speed decreasing at the corner formed by the nominal path at each block transition, a contouring option is used, allowing rounding the corners with an axial tolerance set at $10\, \mu m$. 

The trajectory is run at six different programmed feed rates $F$, geometrically distributed from $1\,000\,mm/min$ to $18\,896\,mm/min$.

\section{Results}

\subsection{Mean value of the norm}

The mean percentage value of the norm along the experimental trajectory, called $\delta_{k,\%}$ for the source $k$, is used to compare the contributions of each source of error and is calculated as follows:

\begin{equation}
\delta_{k,\%} = \frac{1}{n} \sum_{i=1}^{n} \frac{\| ^i \boldsymbol{\delta}_k \|}{ \| ^i \boldsymbol{\delta}_c \| + \| ^i \boldsymbol{\delta}_{qs} \| + \| ^i \boldsymbol{\delta}_d \| } \times 100
\label{eq:delta_k_p}
\end{equation}

\noindent
where $^i \boldsymbol{\delta}_k$ is the $i$-th line of the matrix $\boldsymbol{\delta}_k$, and $k$ can be $c$ for contouring errors, $qs$ for quasi-static geometric errors and $d$ for dynamic errors. Thus, $\delta_{k,\%}$ gives the mean proportion of the total error due to the source $k$ along the trajectory. The repartition of the error sources for the six tests is given in \tablename~1.

For the geometric errors, the distinction is made between the three identified sources. The experiments were carried out with deactivated compensation tables, avoiding any effect of software compensation of the link and motion errors. In an industrial context, the effect of link errors can be compensated, as it is a model-based calculated contribution. 

The increasing relative impact of the thermal drift errors (see \tablename~1) may not be related to the increasing feed rate, but the drift is slowly increasing from test to test while the thermal condition are getting away of the reference thermal state. Moreover, the effect of the thermal drift observed during this series of tests is not reflecting actual machining conditions because the spindle is not running but only regulating its angular position. More sizeable errors can be expected under heavier spindle use.

\tablename~1 also shows the relative importance of the quasi-static geometric errors, compared to the two other sources for the small feed rates. As expected, the summed contributions of the contouring errors and the dynamic errors are increasing from $1.7 \, \%$ to $14.3 \, \%$  for programmed feed rate of $1\,000 \, mm/min$ and $18\,896\, mm/min$ respectively.

\begin{table}[tbh]
\center
\caption{Repartition of the mean value of the error sources along the experimental trajectory for the six tests.}
\label{tab:repartition_contribution}
\vspace{4pt}
\small{
\begin{tabular}{cccccccc}
$F$ & \multirow{2}{*}{$\delta_{c,\%}$}	&&	& $\delta_{qs,\%}$	& 	&&	\multirow{2}{*}{$\delta_{d,\%}$} \\
$(mm/min)$ & 	&&	$\delta_{l,\%}$	& $\delta_{m,\%}$	& $\delta_{td,\%}$	&&	 \\
\hline
$ 1\,000$ & 1.1 && 86.9 & 11.4  & 0.0  && 0.6 \\ 
$ 1\,800$ & 1.7 && 85.9 & 11.3  & 0.4  && 0.7 \\
$ 3\,240$ & 2.4 && 84.7 & 11.1  & 0.9  && 0.9 \\
$ 5\,832$ & 3.9 && 82.6 & 10.8  & 1.1  && 1.5 \\
$10\,498$ & 6.5 && 77.7 & 10.2  & 2.3  && 3.3 \\
$18\,896$ & 9.5 && 72.8 & 9.5  & 3.3  && 4.8 \\
\hline
\end{tabular}}
\end{table}

\begin{table}[t]
\center
\caption{Maximum errors of the three Cartesian components of the geometrical errors contributions in the 3 directions of the machine frame.}
\label{tab:maximum_geometric_errors}
\vspace{4pt}
\small{
\begin{tabular}{p{5.5cm}ccc}
\emph{Contribution}	&	$x$	&	$y$	&	$z$	\\
\hline
Link errors $| \delta_l  |_{max} ~~(\mu m)$ & 4.5	&	54.7	&	16.2	\\
Motion errors $| \delta_m  |_{max} ~~(\mu m)$ & 3.1	&	6.8	&	5.9	\\
Thermal drift $| \delta_{td}  |_{max} ~~(\mu m)$ & 2.0	&	1.2	&	1.8	\\
\end{tabular}}
\end{table}

\begin{table}[h]
\center
\caption{Maximum errors of contouring errors and dynamic errors contributions in the 3 directions of the machine frame for the 6 tested programmed feed rates.}
\label{tab:maximum_errors}
\vspace{4pt}
\small{
\begin{tabular}{ccccccccc}
$F$ &&	\multicolumn{3}{c}{$| \delta_c  |_{max} \, (\mu m)$}&&	\multicolumn{3}{c}{$| \delta_d  |_{max} \, (\mu m)$}\\
$(mm/min)$	&& $x$	&	$y$	&	$z$	&& $x$	&	$y$	&	$z$\\
\hline
$ 1\,000$	&&   8.1 	&  7.5	&	10.1 && 1.1 & 1.7 	& 3.1	\\
$ 1\,800$	&&   7.3   	&  11.2	&	10.3 && 1.5	& 2.9	& 3.2	\\
$ 3\,240$	&&   11.0	&  13.9	&	10.6 && 1.2	& 3.5	& 5.4	\\
$ 5\,832$	&&   11.2	&  16.9	&	10.5 && 1.8	& 4.1	& 7.1	\\
$10\,498$	&&   11.5	&  17.0	&	12.2 && 3.0	& 6.2	& 8.9	\\
$18\,896$	&&   11.5	&  17.0	&	12.5 && 3.7	& 7.6	& 10.4	\\
\end{tabular}}
\end{table}

\begin{table}[h]
\center
\caption{Evolution of the root mean square values of the errors in each direction for different feed rates.}
\label{tab:rms}
\vspace{4pt}
\small{
\begin{tabular}{ccccc}
$F$ &&	\multicolumn{3}{c}{$ \delta_{d,rms} \, (\mu m)$}\\
$(mm/min)$	&& $x$	&	$y$	&	$z$	\\
\hline
$ 1\,000$	&& 0.2 	& 0.2 	& 0.2	\\
$ 1\,800$	0.7 && 0.3	& 0.3	& 0.3	\\
$ 3\,240$	0.9 && 0.4	& 0.4	& 0.5	\\
$ 5\,832$	1.5 && 0.5	& 0.8	& 1.1	\\
$10\,498$	2.2 && 0.8	& 1.9	& 2.1	\\
$18\,896$	2.9 && 1.1	& 2.7	& 3.4	\\
\hline
\end{tabular}}
\end{table}

\begin{figure}[h]
\center
\includegraphics[scale=1]{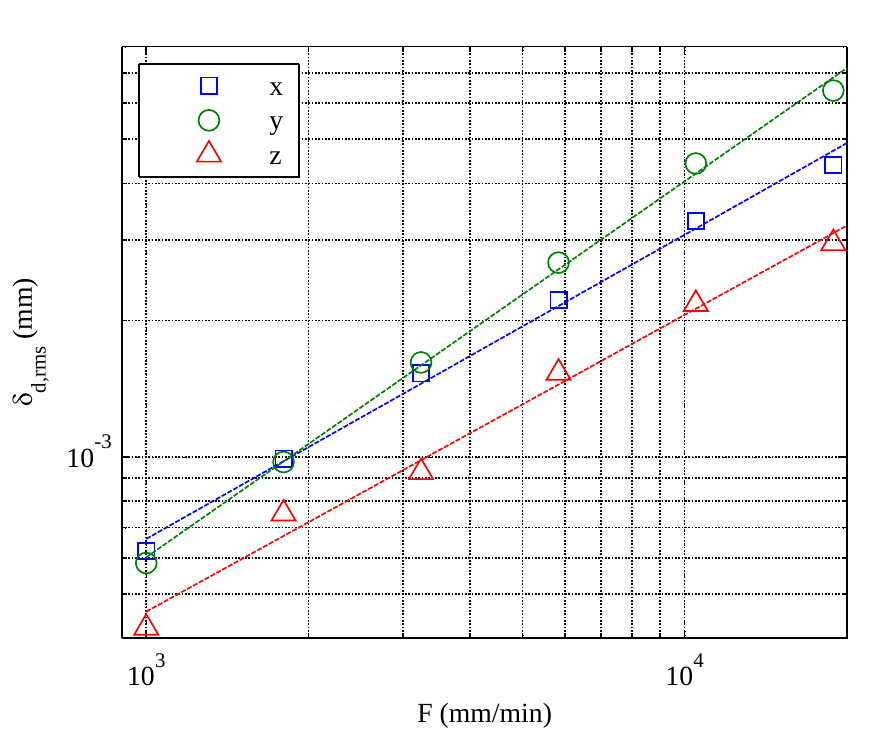}
\caption{Evolution of $ \delta_{d,rms}$ function of the programmed feed rate $F$ in a logarithmic plot. The dashed line are curve fitted logarithmic models.}
\label{fig:rms_deltad}
\end{figure}

\subsection{Maximum errors}

The ISO standards on geometric specifications \cite{iso_gps} generally define tolerance zones in which all the machined surface must be located. Considering this, the most penalising errors are the maximum errors.

The maximum errors $|\delta|_{max}$ of a contribution $\boldsymbol{\delta}$ are defined by the maximum absolute value among the terms of each column $_x\boldsymbol{\delta}$, $_y\boldsymbol{\delta}$ and $_z\boldsymbol{\delta}$ of $\boldsymbol{\delta}$ (eq.\eqref{eq:definition_max}).

\begin{equation}
|\delta|_{max}=\left( \max(|_x\boldsymbol{\delta}|) , \max(|_y\boldsymbol{\delta}|) , \max(|_z\boldsymbol{\delta}|) \right)
\label{eq:definition_max}
\end{equation}

\tablename~2 gives the maximum errors for the quasi-static geometric errors along the experimental trajectory. The maximum drift errors given are the maximum observed during the seven tests. 
Also, the effect of link errors depends on the pose of the machine, and the maximum error can be calculated \emph{offline} for any tool path, and can be higher than the one observed in this study for other points in the working volume of the machine.

\tablename~3 shows the evolution of the contributions of the contouring and dynamic errors with the programmed feed rate. The dynamic errors are not negligible, particularly when the programmed feed rate is over $10\,000\,mm/min$. Even for the first test at $F=1\,000\,mm/min$, dynamic errors occurred, but as shown in \figurename~\ref{fig:essai1}, only locally at the sharp corner of the experimental trajectory. The higher the feed rate, the higher the dynamic errors magnitude. As the feed-rate increases \figurename~\ref{fig:essai6} shows that the dynamic errors also increase.

\subsection{Contributions along the trajectory}

\figurename~\ref{fig:essai1} and \figurename~\ref{fig:essai6} are two examples of the calculated contributions of $\boldsymbol{\delta}_{c}$, $\boldsymbol{\delta}_{qs}$ and $\boldsymbol{\delta}_{d}$ along the experimental trajectory, respectively for the first and the sixth test. The acceleration of each axis is plotted to show that the contouring errors and dynamic errors can, up to a certain point, be related to the acceleration.

The block transition extracted from the NC unit is also plotted. The 17 linearly interpolated segments of the trajectory, previously mentioned in §\ref{par:trajectory}, are visible on those figures. The expected acceleration peaks, which are related to the block transitions, are visible in \figurename~\ref{fig:essai6}. Those acceleration peaks are responsible for the sudden appearance of dynamic loads, themselves related to dynamic errors. In the conditions of those experiments, it shows that the machine geometry is affected by accelerations mainly due to the linear interpolation format. The peak-to-peak magnitude of the resulting dynamic errors is higher than $5\, \mu m$ along the $\boldsymbol{y}$ direction.

{The corner visible near the middle of the trajectory in \figurename~\ref{trajectoire_5x} requires high acceleration of several axes, visible in \figurename~\ref{fig:essai1} and \figurename~\ref{fig:essai6}. The high dynamic load generated on purpose results in visible dynamic geometric errors. Therefore, \figurename~\ref{fig:essai6} shows that the machine accuracy is not equally affected in all directions. For instance, high acceleration of the X-axis seems to have little effect on dynamic errors according to \figurename~\ref{fig:essai6}.}

\subsection{Root mean square of the dynamic errors}

The root mean square value of the dynamic errors along the trajectory in each Cartesian direction are given in \tablename~4. \figurename~\ref{fig:rms_deltad} shows the evolution of $ \delta_{d,rms}$ as a function of $F$ on a logarithmic plot. The dashed lines have equations of the form given by eq.\eqref{eq:model_rms}.

\begin{equation}
\delta_{rms} = \kappa \cdot F^{N}
\label{eq:model_rms}
\end{equation}

The values of $\kappa$ and $N$ are obtained by curve fitting for each error and each direction. The aim of the curve fitting is not to propose a general model for dynamic RMS errors evolution, but only to notice a particular result for the experimental conditions of this work. As the $\kappa$ and $N$ values depend on the experimental trajectory, those results can not lead to general conclusion about the performance of the machine in terms of dynamic stiffness, but it suggests a new performance indicator under dynamic load with a standard trajectory designed for this purpose.

\section{Conclusion}

The high-speed machining context requires high accelerations from the machine. Under those dynamic loads, the machine structure may no longer follow a rigid body behaviour. The new method presented in this paper allows to measure the volumetric error at the tool tip, and to evaluate the contribution of the dynamic geometric errors.

The experiments at high programmed feed rates have been made possible by the use of a non-contact measuring instrument: the CapBall. The CapBall was re-designed to increase its stiffness for more reliable measurement under high dynamic loads.  

The evaluation of the respective magnitude of the contouring errors, the quasi-static geometric errors and the dynamic geometric errors, using the mean values of the norm or the maximum errors, quantifies the relative impact of the dynamic errors, which can reach nearly $5\%$ of the total volumetric error, considering the mean value of the norm.

Furthermore, the study of the root mean square values of the errors shows that in this particular case, the evolution of the RMS errors with respect to the programmed feed rate can be approximated by a logarithmic law.

Finally, the main interest of this experimental work is to propose a method to evaluate dynamic errors directly at the tool tip. Even if the method can only be applied on five -- or more --  axis machine tool, due to the close kinematic chain principle, it can be a powerful mean to validate models for the dynamic behaviour of the structure with \emph{in-situ} measurement. It has been shown that the linear interpolation generates acceleration and dynamic errors peaks. The influence of the NC interpolation and controller command law on dynamic errors can also be investigated with this method.

\section*{Acknowledgements}

This work was partially funded with a Discovery Grant from the National Science and Engineering Research Council of Canada and conducted on equipment purchased with a grant from the Canadian Foundation for Innovation.

\end{document}